\documentclass[12pt]{article}
\usepackage{graphicx}
\begin{document}
\begin{center}

\centerline{\bf Electron Electric Dipole Moment 
induced by Octet-Colored Scalars}
\vskip 1cm
Jae Ho Heo and Wai-Yee Keung
\vskip 1cm
Physics Department, University of Illinois at Chicago, Chicago, IL 60607-7059
\end{center}
\vskip 1cm
\begin{abstract}
  An appended  sector of two octet-colored scalars, each an
electroweak doublet, is an interesting extension of the simple two Higgs
doublet model motivated by the minimal flavor violation. Their
rich CP violating interaction gives rise to a sizable electron
electric dipole moment, besides the quark electric dipole moment 
via the two-loop contribution of Barr-Zee mechanism.
\end{abstract}

\section*{Introduction}
The flavor diversity in the electroweak interaction continues to attract
theoretical curiosity for new physics beyond the standard model. 
Pioneer work of Ref.\cite{Glashow:1976nt} has shown 
a natural mechanism to suppress
the unwanted neutral flavor changing process mediated by the Higgs exchange.
Recent activities address
the general structure of the minimal flavor violation (MFV)
\cite{Manohar:2006ga}. It is noticed 
that octet-colored scalars are able to respect MFV. The general Yukawa
interaction is given by
\begin{equation} 
\mathcal{L}_{Y} = 
-\overline{Q_{L}}\mathbf{y}_{d}
(\phi_{d} + \eta_d O_{d}^{a}T^{a})d_{R}  
-\overline{Q_{L}}\mathbf{y}_{u}
(\widetilde{\phi_{u}}+ \eta_u \widetilde{O_{u}^{a}}T^{a})u_{R}
+ \hbox{ h.c. }
\end{equation}
where $Q_{L}$ refers to the three families of  left-handed quark doublets,
and $d_{R}$ and $u_{R}$  the three families of the up-type and
down-type  right-handed quark singlets. 
The new set of octet-colored scalars 
$O_u$ and $O_d$ couple to quarks in an analogous fashion of the usual
Higgs $\phi_u$ and $\phi_d$ by the Yukawa coupling matrices
${\bf y}_u$ and ${\bf y}_d$ with additional proportional coefficients $\eta_u$
and $\eta_d$, as well as  the color generator $T^{a}$.
All Higgs bosons $\phi_{u,d}$ and $O_{u,d}$ carry the weak hypercharge 
$Y={1\over2}$, in the convention $Q=T_3+Y$.  The usual conjugation
operation $\widetilde{\phi}=i\sigma _{2}\phi^{\ast }$ is adopted.

All  interactions respect a
$Z_{2}$-symmetry, in which  down-type fields $\phi_d$,
$O_d$, $d_R$  are odd, $\phi_{d}\rightarrow
-\phi_{d},O_{d}\rightarrow -O_{d},d_{R}\rightarrow -d_{R}$, but 
up-type fields $\phi_u$, $O_u$, $u_R$  are even. The symmetry is used to
keep  the desirable pattern of interaction that obeys the rule of 
the minimal  flavor violation (MFV).

The complex coefficients $\eta_u$ and $\eta_d$ can produce CP
violating vertices for the 
electric dipole moment (EDM) of the neutron. 
However the
neutron EDM is masked with uncertainty of the
hadronic matrix evaluations. 
It is interesting to ask the question of any contribution to the
electron EDM, reached by ongoing and future experiments\cite{expts}.
It turns out  that one-loop or two-loop contributions to the electron EDM
are absent if there is only one octet-colored scalar electroweak
doublet.
Nonetheless, in this article we show that  with two of these
octet-colored scalars, the electron EDM occurs through an amplitude
$A\to O^+O^-$ of the pseudoscalar neutral Higgs boson $A$ into a pair
of octet-colored charged scalars.  Then, a CP even amplitude for
$A\to\gamma\gamma$ appear even though $A$ is CP-odd. As $A$ and one of the
photon touch down to an electron line, the electron EDM emerges via
Barr-Zee\cite{Barr:1990vd}
mechanism in an overall two-loop diagram, as shown in Fig.~1.
%
\begin{figure}[th!]
\includegraphics[width=6in]{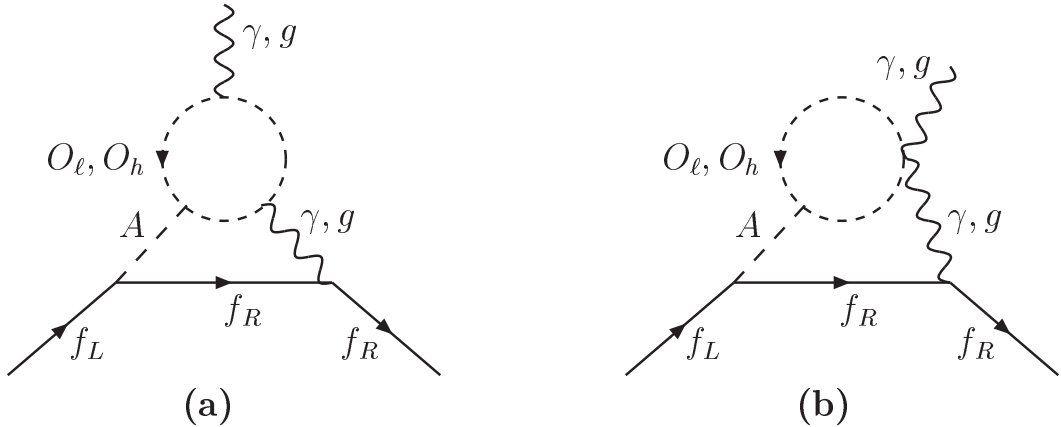}
\caption{Two-loop contributions to EDM and CDM by octet-colored 
  scalars (mirror graphs are not displayed.)}\label{barzee}
\end{figure}
In the past, the electron EDM via Barr-Zee two-loop mechanism has been
exhaustively studied  in scenarios of multi-Higgs 
doublets\cite{Chang:1990sf,Leigh:1990kf}
(without octet-colored scalars) or supersymmetric 
particles\cite{Chang:1998uc,susyedm}.
Our current study  uses  similar  bosonic contribution 
as in \cite{Chang:1998uc,Kao:1992jv}.

We have found that the present limit of the electron EDM has already
restricted the parameter space of CP violation of the octet-colored scalars.

\section*{Higgs potential and $AOO$ coupling}
As we know that if a CP-odd scalar couples to a conjugated pair of
bosons, CP conservation is violated, just like the case in $K_L\to
\pi\pi$.  We are going to find such a coupling in the MFV scenario
with two colorless doublet Higgs $\phi_u$, $\phi_d$, and two
octet-colored Higgs $O_u$, $O_d$.  Rotating the basis of colorless
Higgs,
\begin{equation} 
\left(\begin{array}{c} \phi_1\\ \phi_2\end{array}\right)
  =\left(\begin{array}{cc} \cos\beta & -\sin\beta \\
                           \sin\beta &  \cos\beta     \end{array}\right)
   \left(\begin{array}{c} \phi_u\\ \phi_d\end{array}\right)
\ ,\quad 
   \left(\begin{array}{c} \phi_u\\ \phi_d\end{array}\right)
  =\left(\begin{array}{cc} \cos\beta & \sin\beta \\
                          -\sin\beta & \cos\beta     \end{array}\right)
   \left(\begin{array}{c} \phi_1\\ \phi_2\end{array}\right)  
\ .\end{equation}
we achieve that $\langle\phi_1\rangle=0$ as 
$\tan\beta=\langle\phi_u\rangle/\langle\phi_d\rangle$.
The relevant vertices, allowed by the exact $Z_2$, are
\begin{equation} {\cal L} \supset 
  \eta (\phi_u^\dagger \phi_d) (O_d^\dagger O_u) 
+  \eta' (\phi_d^\dagger \phi_u) (O_d^\dagger O_u) 
+ \ \hbox{ h.c.} 
\label{quartics}
\end{equation}
We start with the first term 
\begin{equation}
\eta (\phi_u^\dagger \phi_d) (O_d^\dagger O_u)
=\eta (\cos\beta\phi_1^\dagger + \sin\beta\phi_2^\dagger)
         (-\sin\beta\phi_1 + \cos\beta\phi_2) O_d^\dagger O_u   
\ .\end{equation}
We single out the pseudoscalar Higgs $A$ from Im($\phi_1$), or
$\phi_1 \to iA/\sqrt{2}$  with $\phi_2\to v/\sqrt{2}$,
\begin{equation}   {\cal L} \supset
\eta (  \cos^2\beta   \phi_1^\dagger\phi_2 
                            - \sin^2\beta   \phi_2^\dagger\phi_1)
                        O_d^\dagger O_u +\hbox{  h.c. } 
\to  -\hbox{$1\over2$}\eta 
  v iA  O_d^\dagger O_u  +\hbox{  h.c. } 
\end{equation}
\begin{equation} {\cal L} \supset    - i \hbox{$1\over2$}
v  A  ( \eta O_d^\dagger O_u - \eta^*O_u^\dagger O_d) 
=
i  \hbox{$1\over2$}  v  A
( \ O_u^\dagger \ ,\ O_d^\dagger )
\left(\begin{array}{cc} 0 & \eta^* \\
                     - \eta & 0\end{array}\right)
\left(\begin{array}{c}   O_u    \\  O_d \end{array}\right)  
\ .\end{equation}
Including the $\eta'$ coupling,
we  combine results and obtain
\begin{equation} {\cal L} \supset    
i  \hbox{$1\over2$}   v  A
( \ O_u^\dagger \ ,\ O_d^\dagger )
\left(\begin{array}{cc} 0 & \eta^*-{\eta'}^* \\
                        \eta'-\eta & 0\end{array}\right)
\left(\begin{array}{c}   O_u    \\  O_d \end{array}\right)  
\ .\end{equation}
Now We look at the charged components of the color octets.
The light and heavy  mass states   $O_\ell$ and $O_h$  are given by
the diagonalization of the mass matrix
\begin{equation} {\cal L} \supset - 
( \ O_u^\dagger \ ,\ O_d^\dagger )
\left(\begin{array}{cc} M^2_{\ uu} & M^2_{\ ud} \\
                        M^2_{\ du} & M^2_{\ dd}\end{array}\right)
\left(\begin{array}{c}   O_u    \\  O_d \end{array}\right)  
\ ,\end{equation}
where  $M^2_{\ uu}$ and $ M^2_{\ dd}$ are real, and
the complex off-diagonal mass squared  $M^2_{\ du}=|M^2_{\ du}|e^{i\Delta}$, 
\begin{equation} -M^2_{du}= 
(\eta'+\eta) v^2\sin\beta\cos\beta   
\ .\end{equation}
We can absorb the phase $\xi$ of   $(\eta'-\eta) \equiv
|\eta'-\eta| e^{i\xi}$ into the field $O_u$.  Then $AOO$ coupling
matrix becomes real and antisymmetric, but 
the remaining phase $\Delta-\xi$ in the off-diagonal mass squared is
genuine and not removable.  The unitary diagonalization transformation is
usually complex,
\begin{equation} 
   \left(\begin{array}{c} O_u\\ O_d\end{array}\right)
  =\left(\begin{array}{cc} \cos\psi & \sin\psi e^{i\delta} \\
               -\sin\psi e^{-i\delta}  & \cos\psi     \end{array}\right)
   \left(\begin{array}{c} O_\ell \\ O_h\end{array}\right)   
\ ,\end{equation}
where the $2\times2$ unitary matrix is denoted as $U$.
\begin{equation} U^\dagger \left(\begin{array}{cc} 0 & -1 \\
                        1 & 0\end{array}\right) U
=\left(\begin{array}{cc} -i\sin\delta \sin2\psi & \cdots \\
                        \cdots &     i\sin\delta \sin2\psi \end{array}\right)
\ .\end{equation}
Note that diagonal entries arise due to 
the CP violating phase $\delta=\xi-\Delta$.  Traceless  is a consistent check.
We denote $|\eta' -\eta| \equiv \lambda$.
The relevant  diagonal $AOO$ interaction becomes
\begin{equation} {\cal L} \supset
\hbox{$1\over2$}\lambda   v  \sin2\psi \sin\delta \ A  
(O_\ell^\dagger O_\ell - O_h^\dagger O_h)  
\ .\end{equation}
The interaction can produce the electron EDM via Barr-Zee mechanism
with the charged octet Higgs $O$ in the inner loop.
Also note that if  masses of  eigenstates $O_\ell$ and $O_h$
become equal, the CP violating effect disappears.
Other quartic couplings not given in Eq.(\ref{quartics}) do not generate the
interaction between $A$ and the $O$ pair.

As the neutral pseudoscalar $A$ is not constrained very much from LEP
data, we can choose a light value, for example $m_A \sim M_Z$ about
100 GeV in our numerical analysis. The four degrees of freedom in the
two real $M^2_{\ uu}$ and $M^2_{\ dd}$ and the complex $M^2_{\ du}$
can be replaced by the other four parameters, two masses $m_{O_\ell}$
and $m_{O_h}$ of the light and heavy charged octet Higgs, and two
angles $\psi$ and $\delta$.

\section*{Barr-Zee Amplitude}
The CP violating short distance physics induces the effective vertex
of the electric dipole $d_f$ of the fermion $f$
coupled with the electromagnetic field strength $F_{\mu\nu}$,
\begin{equation} {\cal L}_{\rm eff} \supset -\hbox{$i\over2$} \ 
d_f \ (\bar f \sigma^{\mu\nu}\gamma_5 f )\
F_{\mu\nu} 
\ .\end{equation}
The two-loop Barr-Zee diagram also involves the 
pseudoscalar  coupling of $A$  to fermions,
\begin{equation} {\cal L} \supset 
- (1/v) 
[\tan\beta ( m_e\bar e i\gamma_5 e  + m_d \bar d i\gamma_5 d)
+\cot\beta\ m_u  \bar u i\gamma_5 u ] A 
\ ,\end{equation}
where the electron shares similar pattern of the Yukawa coupling as the
down quark.
Evaluating the Barr-Zee diagram, we obtain
\begin{equation}
\left({d_e\over e}\right)
=-{\alpha \lambda  m_e \over 8\pi^3 m_A^2 } \tan\beta\sin(2\psi)\sin\delta  
\left[ F\left({m_{O_\ell}^2 \over m_A^2}\right) 
  -F\left({m_{O_h}^2 \over m_A^2}\right)  \right] 
\ .\end{equation}
Eight color channels of $O$ have been summed.  The relevant
function is defined by
\begin{equation} F(z) =\int_0^1 {x(1-x)\over z-x(1-x)}\log {x(1-x)\over z}  
\ ,\end{equation}
which has asymptotic behaviors,
\begin{equation}  F(z)\longrightarrow 
\left\{
\begin{array}{c c}
{-0.344}   & \hbox{ as } z=1\ , \\
-{1\over6z}\ln z -{5\over 18z}  & \hbox{ for } z\gg 1\ ,\\
(2+\ln z)                       & \hbox{ for } z\ll 1\ .\end{array}
\right.  
\ .\end{equation}
The electron EDM is evaluated as the difference 
between two contributions from the
light and  heavy color octets $O_\ell$ and $O_h$. 
We  illustrate  in Fig.~2 the EDM component 
due to the light $O_\ell$ pretending 
$m_{O_h}\to \infty$ for given $\psi$ and $\delta$.
The electron EDM contribution from the light octet is plotted 
versus $m_A$ for various 
$m_{O}=300, \ 500,  \ 800,\ 1000$ GeV,
assuming $\lambda\tan\beta\sin(2\psi)\sin\delta=1$.
The current experimental constraint\cite{Regan:2002ta}
is shown by the horizontal double-line, which has already imposed a 
constraint on the parameter space of the model.
\begin{figure}[th!]
\includegraphics[width=6in]{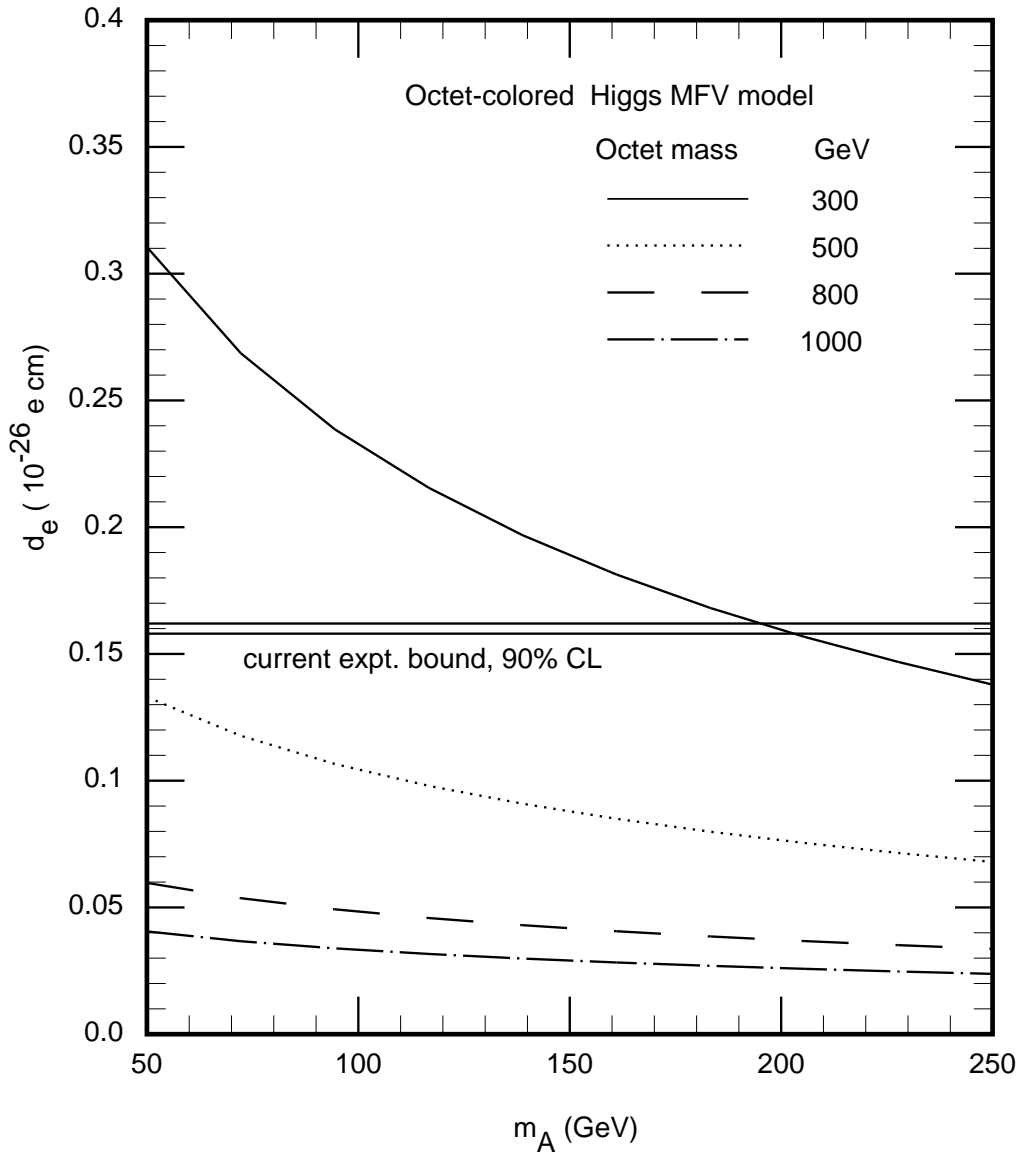}
\caption{ \small
\label{numedm}
Predicted electron EDM versus $m_A$ for various $m_{O}=$
300 (solid), 500 (dotted), 800 (dashed), 1000 (dashed-dotted) GeV,
with  $\lambda\tan\beta\sin(2\psi)\sin\delta=1$.
The horizontal double-line is the present experimental upper bound.
}
\end{figure}

Similarly, we can obtain the  quark EDM  by rescaling parameters,
\begin{equation}
\left({d_{d,s}\over d_e}\right)
=\left({-{1\over 3}\over -1}\right)
\left({ m_{d,s} \over m_e}\right)
\ ,\qquad
\left({d_{u}\over d_e}\right)
={1\over \tan^2\beta}\left({{2\over 3}\over -1}\right)
\left({ m_{u} \over m_e}\right)
\ .\end{equation} 
We can also work out\cite{Chang:1990twa} 
the associated color  dipole moment
(CDM) $d^C_q$ of the quark $q$  in the effective vertex
by  replacing
color factors and couplings.
On the other hand, Ref.\cite{Manohar:2006ga}, has studied the
contribution of the color octet to 
the gluon dipole\cite{Braaten:1990zt}, 
known as the Weinberg operator\cite{Weinberg:1989dx}, which 
is subjected to large renormalization 
suppression\cite{Braaten:1990gq,Boyd:1990bx,Chang:1990jv}.
Due to the sophisticated hadron physics, 
we only have a qualitative relation  between the neutron EDM
and the CP violating coefficients, mainly based on naive dimensional 
analysis\cite{Manohar:1983md}.
On the contrary, the electron EDM is well predicted because it is 
independent of the hadronic structure.

\section*{Conclusion}
The octet-colored scalars if exist will be copiously 
produced\cite{Manohar:2006ga,Gerbush:2007fe,Gresham:2007ri,Dobrescu:2007yp}
by LHC in coming years because of its strong interaction and its larger color
representation. When  their masses and couplings are determined, the
electron EDM measurement is the key to disclose its nature of the CP violation.
We show in this paper that the electron EDM measurement is sensitive
to the CP violating phase in the octet-colored scalars.

This work was supported in part by U.S. Department of Energy under
grant number DE-FG02-84ER40173.


\begin{thebibliography}{xx}
%
%
\bibitem{Glashow:1976nt}
  S.~L.~Glashow and S.~Weinberg,
  Phys.\ Rev.\  D {\bf 15}, 1958 (1977).
\bibitem{Manohar:2006ga}
  A.~V.~Manohar and M.~B.~Wise,
  Phys.\ Rev.\  D {\bf 74}, 035009 (2006)
  [arXiv:hep-ph/0606172].
%
\bibitem{expts}
  D.~Kawall, F.~Bay, S.~Bickman, Y.~Jiang and D.~DeMille,
  AIP Conf.\ Proc.\  {\bf 698},  192 (2004);
  D.~Kawall, F.~Bay, S.~Bickman, Y.~Jiang and D.~DeMille,
  Phys.\ Rev.\ Lett.\  {\bf 92}, 133007 (2004)
  [arXiv:hep-ex/0309079].
%
\bibitem{Barr:1990vd}
  S.~M.~Barr and A.~Zee,
  Phys.\ Rev.\ Lett.\  {\bf 65}, 21 (1990)
  [Erratum-ibid.\  {\bf 65}, 2920 (1990)].
%
\bibitem{Chang:1990sf}
  D.~Chang, W.-~Y.~Keung and T.~C.~Yuan,
 Phys.\ Rev.\  D {\bf 43}, 14 (1991).
%
\bibitem{Leigh:1990kf}
  R.~G.~Leigh, S.~Paban and R.~M.~Xu,
  Nucl.\ Phys.\  B {\bf 352}, 45 (1991).
%
\bibitem{Chang:1998uc}
  D.~Chang, W.-~Y.~Keung and A.~Pilaftsis,
  Phys.\ Rev.\ Lett.\  {\bf 82}, 900 (1999)
  [Erratum-ibid.\  {\bf 83}, 3972 (1999)]
  [arXiv:hep-ph/9811202].
%
\bibitem{susyedm}
  D.~Chang, W.~F.~Chang and W.-~Y.~Keung,
  Phys.\ Lett.\  B {\bf 478}, 239 (2000)
  [arXiv:hep-ph/9910465];
%
  A.~Pilaftsis,
  Phys.\ Lett.\  B {\bf 471}, 174 (1999)
  [arXiv:hep-ph/9909485];
%
  D.~Chang, W.~F.~Chang and W.-~Y.~Keung,
  Phys.\ Rev.\  D {\bf 66}, 116008 (2002)
  [arXiv:hep-ph/0205084].
%
\bibitem{Kao:1992jv}
  C.~Kao and R.~M.~Xu,
  Phys.\ Lett.\  B {\bf 296}, 435 (1992).
%
%
\bibitem{Regan:2002ta}
  B.~C.~Regan, E.~D.~Commins, C.~J.~Schmidt and D.~DeMille,
  Phys.\ Rev.\ Lett.\  {\bf 88} (2002) 071805.
%
\bibitem{Chang:1990twa}
  D.~Chang, W.-~Y.~Keung and T.~C.~Yuan,
  Phys.\ Lett.\  B {\bf 251}, 608 (1990).
%
\bibitem{Braaten:1990zt}
  E.~Braaten, C.~S.~Li and T.~C.~Yuan,
  Phys.\ Rev.\  D {\bf 42}, 276 (1990).
%
\bibitem{Weinberg:1989dx}
  S.~Weinberg,
  Phys.\ Rev.\ Lett.\  {\bf 63}, 2333 (1989).
%
\bibitem{Braaten:1990gq}
  E.~Braaten, C.~S.~Li and T.~C.~Yuan,
  Phys.\ Rev.\ Lett.\  {\bf 64}, 1709 (1990).
%
\bibitem{Boyd:1990bx}
  G.~Boyd, A.~K.~Gupta, S.~P.~Trivedi and M.~B.~Wise,
  Phys.\ Lett.\  B {\bf 241}, 584 (1990).
%
\bibitem{Chang:1990jv}
  D.~Chang, W.-~Y.~Keung, C.~S.~Li and T.~C.~Yuan,
  Phys.\ Lett.\  B {\bf 241}, 589 (1990).
%
\bibitem{Manohar:1983md}
  A.~Manohar and H.~Georgi,
  Nucl.\ Phys.\  B {\bf 234}, 189 (1984).
\bibitem{Gerbush:2007fe}
  M.~Gerbush, T.~J.~Khoo, D.~J.~Phalen, A.~Pierce and D.~Tucker-Smith,\\
  arXiv:0710.3133 [hep-ph].
\bibitem{Gresham:2007ri}
  M.~I.~Gresham and M.~B.~Wise,
  Phys.\ Rev.\  D {\bf 76}, 075003 (2007)
  [arXiv:0706.0909 [hep-ph]].
%
\bibitem{Dobrescu:2007yp}
  B.~A.~Dobrescu, K.~Kong and R.~Mahbubani,
  arXiv:0709.2378 [hep-ph].
\end{thebibliography}
\end{document}